\documentclass[12pt,a4paper]{article}
\begin{document}
\newcommand{\eq}[1]{~(\ref{#1})}
\newcommand{\BEQ}{\begin{equation}}
\newcommand{\EEQ}{\end{equation}}
\newcommand{\norm}[1]{\mbox{$\left| \left| #1 \right| \right| $}}
\newcommand{\abs}[1]{\mbox{$\left| #1 \right| $}} 
\newcommand{\bra}[1]{\mbox{$\langle \left. #1 \right| $}}
 \newcommand{\ket}[1]{\mbox{$\left|  #1 \rangle \right. $}}
\newcommand{\DEG}[1]{\mbox{$ #1^{\rm o}$}}
\newcommand{\lappr}{\mbox{$\stackrel{<}{\sim}$}} 
\newcommand{\gappr}{\mbox{$\stackrel{>}{\sim}$}} 
\newcommand{\mr}[1]{\mbox{\rm #1}} 
\newcommand{\ETC}{\mbox{\em etc.\/ }}
\newcommand{\VIZ}{\mbox{\em viz.\/ }}
\newcommand{\CF}{\mbox{\em cf.\/ }}
\newcommand{\IE}{\mbox{\em i.e. \/}}
\newcommand{\ETAL}{\mbox{\em et. al.\/ }}
\newcommand{\EG}{\mbox{\em e.g.\/ }}
%
%
\begin{flushright}
DFF-274/03/1997 (Florence)\\
JHU--TIPAC--97006 (Johns Hopkins)\\
March  1997 
\end{flushright}
\vspace*{8mm}
\begin{center}
{\Large\bf Behavior of Neutrinos in Stochastic Magnetic Fields}\\[5truemm]
G. Domokos and S. Kovesi--Domokos\\[2mm]
Dipartimento di Fisica, Universit\'{a} di Firenze\\
Florence, Italy\\[1mm]
and\\[1mm]
The Henry A. Rowland Department of Physics and Astronomy\\
The Johns Hopkins University\footnote{Permanent address. 
E--mail:~SKD@HAAR.PHA.JHU.EDU}\\
Baltimore, MD 21218
\end{center}
\vspace*{4mm}
\begin{quote}
If massive neutrinos possess magnetic moments, they can undergo spin
flip in a magnetic field. The magnetic fields needed for a meaningful
measurement of neutrino moments could be very high and may
occur in astronomical objects such as some supernovae or active
galactic nuclei: they are typically chaotic ones. We develop the
general theory of the passage of  neutrinos through such fields. 
We also develop a simple model which becomes solvable in the
high energy limit. Both helicities occur with equal probability,
independently of the initial distribution. Observational consequences
are discussed.
\end{quote}
\vspace*{5mm}
\begin{flushleft}
\end{flushleft}
\section{Introduction}
The existence and magnitude of magnetic moments of neutrinos is one of
the important outstanding questions of particle physics. It is well
known that in the standard model of electroweak interactions neutrinos
are purely left handed; hence, they are massless and cannot possess
magnetic moments either. However, in some minimal extensions of the
standard model and also in in some grand unified models, \EG in the
SO(10) model, there are both left and right handed neutrinos present.
Therefore, there is the possibility giving them mass and also, 
chirality flipping (Pauli type) interactions with the electromagnetic
field. In such models, higher order electroweak interactions give rise
to anomalous magnetic moments {\em via} loops in which a W and a
charged lepton is circulating. This was first noticed by Lee and
Shrock~\cite{leeshrock} a long time ago. The magnetic moment thus 
arising, however, is a very small one, due, mainly, to the large mass
of the W boson. It is given by the expression,
\[
\mu_{\nu} = \frac{3 G_{F} e m_{\nu}}{8\sqrt{2} \pi^{2}}
\approx 3\times 10^{-19}\frac{m_{\nu}}{1{\rm eV} \mu_{B}}.
\]
Here $G_{F}$ stands for the Fermi coupling constant and $\mu_{B}$
is the Bohr magneton.

There exist fairly reliable upper limits on the masses of neutrinos;
using those, the magnitudes of the moments predicted by this
formula come out to be  several orders of magnitude lower than the current
observational
upper limits, see, \EG \cite{kimpevsner} for a recent review. 
(Current limits are typically around $10^{-12}\mu_{B}$ for 
flavor diagonal moments and a few orders of magnitude higher
for transition moments.)  As a
consequence, any observation of a neutrino magnetic moment indicates
the presence of some significant physics beyond the standard model.
Using the observational upper limits on the moments,  however,one concludes
that such a measurement is likely to take place by means of a neutrino
telescope. Indeed, the characteristic length associated with a 
magnetic dipole moment of magnitude $\mu$ placed in a magnetic field
of  strength 
$B$ is given by $L = 1/\mu B$. (Roughly speaking, this is the
distance over which helicity flip occurs with a substantial
probability.) In convenient
units,
\[
\frac{L}{km} \approx 3\times 10^{-2}\frac{\mu_{B}{\rm Gauss}}{\mu B}.
\]
For a magnetic moment of $10^{-12}\mu_{B}$ moving even 
in a field of 
1~Megagauss, the characteristic length is about $3\times 10^{4}$km.
Clearly, no terrestrial experiment can be designed for such a
measurement. By contrast, an active galactic nucleus (AGN) or a
supernova,
to quote but two examples,  provides
us with the right laboratory for measuring small magnetic moments
either because of the high magnetic fields and/or by the large size of the
object in question.

The trouble is that magnetic fields in astronomical objects are
generally
chaotic ones and the description of the motion  of a magnetized
neutrino in such an environment is hard to describe by the methods 
particle physicists are used to. The main  purpose of the present work is
to
develop the general formalism necessary for solving such a problem.
 Section~\ref{general} is devoted to the development of this formalism,
while in the subsequent one (Sec.~\ref{nullplane}) 
we describe the relevant  dynamics in front form. 
A solvable  model (involving a reduced
number of neutrino flavors) is described in Section~\ref{solvable}.  
Finally, the results and their observational
consequences are discussed in Sec.~\ref{discussion}.  
\section{General Formalism}
\label{general}
The problem of investigating the behavior of a microscopic system
in a random environment is often faced in condensed matter physics.
The classic example is, of course, Brownian motion, but similar
problems occur in the statistical theory of turbulence and many 
other situations as well. 
Typically, such systems are modeled by an equation of motion of the
form:
\BEQ
\partial_{t}X + F[X] = f,
\label{oldrandom}
\EEQ
where $X$ is an element of a vector space (the phase space of dynamical
variables), the functional $F[.] $ is often, but not necessarily, local and
time independent. Finally, $f$ represents a random force perturbing
the otherwise deterministic system.

The modern theory of such systems was developed by Martin, Siggia and 
Rose~\cite{martin} using a canonical formalism and by De~Dominicis and
Peliti~\cite{dedominicis} using a functional integral formalism.
For the purpose of generalization to a relativistic system, the formalism
developed in ref.~\cite{dedominicis} is more suitable.

There is an important difference, however, between  systems described by
an equation like \eq{oldrandom} and ones we are interested in. In 
\eq{oldrandom} the random perturbing force appears as an inhomogeneous
term on the right hand side of the otherwise deterministic dynamical
equation, whereas for a particle in a random 
magnetic field, the latter
appears typically in the form ${\bf B}\cdot {\bf \Sigma}\varphi$, where 
${\bf \Sigma} $ is a spin operator and $\varphi$ is an amplitude describing
the propagation of the particle\footnote{In this work we are
interested in neutral particles, hence interaction terms proportional to the
electric charge are absent.}. Hence, the formalism has to be modified 
in order to take this circumstance into account. 

We use a Hamiltonian formalism; manifest covariance is of little value, since 
the rest frame of the astronomical object provides us with a preferred frame
of reference. The Hamiltonian is taken to be of the form:
\BEQ
H = H_{0} + \mu {\bf M}\cdot {\bf B},
\label{hamiltonian}
\EEQ
where $H_{0}$ governs the propagation of the particle in the absence 
of a magnetic field. The quantity ${\bf M}$ describes the
coupling of the particle to the magnetic field and it may 
contain degrees of freedom other than just spin: \EG various
flavors a particle comes in may have different types of magnetic
couplings. Correspondingly, $\varphi$ is a vector in the space of 
helicities and
flavors.
Finally, $\mu$ is the magnitude of some average of
the magnetic moments involved: it carries the necessary dimension and 
the absence of a magnetic coupling can be described as the limit 
$\mu \rightarrow 0$.

The equation governing the propagation of the particle thus reads:
\BEQ
\partial_{t}\varphi + H_{0}\varphi + \mu {\bf M}\cdot {\bf B}\varphi =0
\label{dynamical}
\EEQ
This equation is a deterministic one: there is no randomness involved.
Assuming that one finds a solution of \eq{dynamical} for
an {\em arbitrary} magnetic field and one is given
the distribution of ${\bf B}$ over the ensemble of the magnetic fields of
interest,  one can determine the average behavior of the solution as well
as the fluctuations around the average.

Formally, let $\varphi \left(x |{\bf B}\right)$ denote the solution of
\eq{dynamical} as a functional of the magnetic field, where $x$ stands for
the spatial coordinates and the time. Assume further 
that one is given the
characteristic functional, $Z\left[ {\bf j} \right]$ of the distribution of the
magnetic fields over the ensemble. Then the average behavior of 
$\varphi$ is given by the expression:
\BEQ
\langle \varphi \left( x |{\bf B}\right)\rangle = \varphi \left( x \left|
\right. \frac{-i \delta}{\delta {\bf j}}\right) Z\left[{\bf j}\right]_{j=0}.
\label{average}
\EEQ 
The fluctuations around the average are given by similar expressions 
involving higher functional derivatives.

Equation \eq{average} can be useful provided two things are specified:
\begin{itemize}
\item The characteristic functional, $ Z\left[ {\bf j}\right]$
\item The solution of equation \eq{dynamical} for an ``arbitrary''
magnetic field.
\end{itemize}

It is clear that, for any system but the very simplest ones, the two items
specified above cannot be given in a closed form. (For a simple system to
be discussed in Section~\ref{solvable}, one can give analytic expressions in
the limit of high energies.) What one needs instead is a meaningful
approximation scheme permitting the calculation of the quantities needed.

First of all, we notice that for any physically acceptable ensemble of 
magnetic fields, the characteristic functional can be written as a
functional integral, \VIZ
\BEQ
Z\left[ {\bf j}\right] = \int {\cal D}{\bf B} \exp - S\left[{\bf B}\right]
\times \exp i\langle {\bf j}\cdot {\bf B}\rangle .
\label{characteristic}
\EEQ
Here $S$ stands for a generalized entropy functional and $\langle \ldots
\rangle$ denotes integration over continuous variables and summation over 
discrete ones. The functional measure is normalized in such a way that
$Z\left[ 0 \right] =1$. The generalized entropy has two important
properties:
\begin{enumerate}
\item $S\geq 0$
\item $\partial_{i}\frac{\delta}{\delta j_{i}} Z\left[ {\bf j}\right] =0
$.
\end{enumerate}
The first property follows from the fact that $Z$ is the characteristic
functional of a probability distribution, the second one means that the
magnetic field is solenoidal, as it should be. In general,
the entropy functional is determined from solving Maxwell's equations
for some random current distribution; in the next section we shall
consider a simple case of a static field. Thereafter, one has to decide on
physical grounds the expression of the entropy functional.

The simplest, physically reasonable approximation is to consider a
Gaussian distribution,
\BEQ
S = \frac{1}{2}\langle {\bf B} C^{-1} {\bf B}\rangle,
\label{Gaussian}
\EEQ
where $C$ is the correlation operator. In what follows, we shall assume
a distribution of this type.

Next, we construct the generating functional for the moments of
$\varphi$. We have to average over the manifold of solutions of
equation \eq{dynamical}, so we have:
\begin{eqnarray}
Z[u]& = &\int {\cal D}B {\cal D}\varphi 
\exp -\frac{1}{2}\langle B C^{-1} B\rangle
 \exp i \langle u \varphi\rangle \nonumber \\
    & \times & \delta \left[ i \partial_{t}\varphi + H_{0}\varphi + \mu 
{\bf B}\cdot {\bf M}\varphi\right]  Det \left( i\partial_{t}
+ H_{0} + \mu {\bf B}\cdot {\bf M}\right),
\label{thegenerator}
\end{eqnarray}
where $u$ is the source of $\varphi$. The determinant is inserted in
order to get the correct measure on the manifold of solutions. One can use
the integral representation of the $\delta$-functional, 
$\delta[.]$  in order to cast \eq{thegenerator} into a form to which 
techniques familiar from quantum field theory are directly applicable.
We write:
\BEQ
\delta \left[i \partial_{t}\varphi + H_{0}\varphi + \mu 
{\bf B}\cdot {\bf M}\varphi \right]  = \int {\cal D}\varphi^{\dag}
\exp i \langle \varphi^{\dag}\left( i \partial_{t}\varphi +
H_{0}\varphi + 
\mu  {\bf B}\cdot {\bf M}\varphi \right)\rangle
\label{deltafunctional}
\EEQ
It is to be noted that in \eq{deltafunctional}, $\varphi^{\dag}$ is 
not the Hermitean conjugate of $\varphi$, but an independent functional
argument; the notation is used because of the transformation properties 
of $\varphi^{\dag}$ under any symmetry group the argument of the delta--
functional may have.
Using \eq{deltafunctional} the generating functional  for the moments of
$\varphi$ can be brought to a form familiar from quantum field theory,
\VIZ
\begin{eqnarray}
Z\left[u, u^{\dag}\right] & = & \int {\cal D}B {\cal D}\varphi {\cal D}
\varphi^{\dag} \exp -\frac{1}{2}\langle {\bf B}C^{-1}{\bf B} \rangle 
\nonumber \\
   & \times & \exp i\langle \left(\varphi^{\dag}u  + u^{\dag}\varphi
\right) \rangle
\nonumber \\
   & \times & \exp i \langle \varphi^{\dag}
\left( i \partial_{t}\varphi + H_{0}\varphi + \mu 
{\bf B}\cdot {\bf M}\varphi \right)\rangle \nonumber \\
   & \times & Det \left( i\partial_{t}
+ H_{0} + \mu {\bf B}\cdot {\bf M}\right).
\label{fieldtheory}
\end{eqnarray} 
(We introduced a source for $\varphi^{\dag}$ as well, in order to be able to
generate Green's functions.)

Equation \eq{fieldtheory} is recognized as being formally
equivalent to  a quantum field theory for
the variables $\varphi$ and $\varphi^{\dag}$, the variable ${\bf B}$ merely
serving as a mediator of interactions. The functional determinant can 
be handled by familiar techniques, typically, by introducing Fadeev--Popov
ghost fields. The Gaussian integration over ${\bf B}$ can be carried out
explicitly; however, this may not be the most desirable form of 
\eq{fieldtheory}.

We conclude that the problem of describing the propagation of a neutral 
particle (a neutrino in particular) in a random magnetic field 
can be handled in a way familiar in quantum field theory. Familiar
approximation methods, such as loop expansions, \ETC are readily
applicable to this problem as they are to any quantum field theory.

So far, we have not specified the form of the dynamics entering
the preceding equations. This is an important question:
casting the dynamics into an appropriate form leads to significant
simplifications.
\section{Dynamics in Front Form.}
\label{nullplane}

We use the front form of
dynamics~\cite{dirac}. As explained in a previous 
work~\cite{heneutr}, this
formulation of dynamics is advantageous in a situation in which one
considers the propagation of high energy particles ($E\gg m$, where
$m$ is the rest mass) and in which certain discrete symmetries, such as
$C$ and $P$ play no significant role. Clearly, the propagation of high
energy neutrinos falls into this category. In what follows, we
assume an arbitrary flavor structure: all physical observables
(masses, magnetic moments, \ETC) are matrices in flavor space. 
However, flavor indices need  not be explicitly exhibited.

We begin with the usual Dirac Lagrangian of a particle in an external 
electromagnetic field, $F_{\mu \nu}$:
\BEQ \label{diraclagrangian}
L = \overline{\psi}\left( \gamma^{\mu}\partial_{\mu} + m 
+ \frac{1}{2}\mu F^{\mu \nu}\sigma_{\mu \nu}\right)\psi
\EEQ
We work in the rest frame of the magnetic field.  Assuming the 
field to be a static one\footnote{In physical terms,
this means that the characteristic time scale of change of  the field is
large compared to the time of passage of neutrinos. It is
a straight forward matter to generalize the formalism for arbitrary
electromagnetic fields.}, 
we can set $F_{0i}=0$, $F_{ij}=\epsilon_{ijk}B_{k}$.
In the case of 
interest one has to solve the Dirac equation in an {\em arbitrary}
static magnetic field, since we want to average the solution over an
ensemble of  the $B_{i}$. No explicit solution is 
known for such a problem. However, we proceed to show that in the
{\em high energy limit} the problem can be solved in a closed form.

We introduce a coordinate system in which two of the coordinates
are null directions corresponding to characteristic lines of a relativistic 
wave equation, \VIZ:
\BEQ
t=\frac{1}{\sqrt{2}}\left( x^{0} - x^{3}\right), z=\frac{1}{\sqrt{2}}\left(
x^{0} + x^{3}\right) \quad {\rm and} \quad x^{A}; \quad (A=1,2).
\label{nullcoordinates}
\EEQ
Correspondingly, the metric is of the form,
\BEQ
g_{zt}=g_{tz}=1, \quad g_{AB}= -\delta_{AB},
\label{metric}
\EEQ
and all other components vanish.
The momentum components conjugate to $z$ and $t$ are,
\BEQ
k \sim p_{3}\sqrt{2},\quad h\sim \frac{m^{2}+\vec{p}^{2}}{2k}\qquad
(p_{3}\rightarrow \infty, \quad \vec{p}\quad {\rm finite}),
\EEQ
respectively.
In this equation, $\vec{p}$ is the momentum transverse to $p_{3}$.

A Dirac spinor can be decomposed along the null directions 
given in \eq{nullcoordinates} by introducing
the mutually orthogonal projectors,
\BEQ P_{t}=\frac{1}{2}\gamma_{t}\gamma^{t}, \quad P_{z}= \frac{1}{2}
\gamma_{z}\gamma^{z}
\label{nullprojectors}
\EEQ
In what follows, we use the shorthand,
\BEQ
\phi = P_{t}\psi, \quad \chi = P_{z}\psi
\label{shortnames}
\EEQ

It is a straightforward matter to decompose \eq{diraclagrangian}
according to the conjugate null directions and express it in terms of
the variables $\phi$ and $\chi$. The purpose of such an exercise is a very 
simple one. If, for the sake of definiteness, $t$ is regarded the ''time''
variable describing the dynamics of the system, only $\phi$ obeys
an equation containing $\partial_{t}$. Hence, the conjugate component
of the Dirac spinor obeys only an equation of constraint. The constraint 
can be, in turn, solved before one attempts to attack the problem of dynamics.

After carrying out the decomposition of \eq{diraclagrangian} according
to the null directions, one finds: 
\begin{eqnarray}
L &=& \sqrt{2}\left[ \phi^{\dag}\left( i \partial_{t} -i \sqrt{2} \mu
\epsilon^{AB}\gamma_{A}B_{B}\right) \phi \right. \nonumber \\
  &+& \left. \chi^{\dag}\left( i\partial_{z} - i\sqrt{2} \mu
\epsilon^{AB} \gamma_{A}B_{B}\right) \chi \right] \nonumber \\
  &+& \frac{1}{\sqrt{2}}\left[ \phi^{\dag}\gamma^{z}\left(
i\gamma^{A}\partial_{A} +m - \frac{i}{\sqrt{2}}\mu B_{3}\,\epsilon_{AB}
\gamma ^{A} \gamma^{B}\right)\chi \right. \nonumber \\
  &+& \left.  \chi^{\dag}\gamma^{t}\left( i \gamma^{A}\partial_{A} + m 
 - \frac{i}{\sqrt{2}}\mu B_{3}\,\epsilon_{AB}
\gamma^{A}\gamma^{B}\right)\phi \right]
\label{decomposedlagrangian}
\end{eqnarray}

Variation of \eq{decomposedlagrangian} with respect to 
$\chi^{\dag}$ gives the constraint:
\BEQ
\left( i\sqrt{2}\partial_{z} - 2i \mu
\epsilon^{AB}\gamma_{A}B_{B}\right)\chi
+ \frac{1}{\sqrt{2}}\gamma^{t}\left( i\gamma^{A}\partial_{A}
+\frac{i}{\sqrt{2}}\mu \epsilon_{AB}\gamma^{A}\gamma^{B}B_{3} +m\right)\phi
=0.\label{constraint}
\EEQ
The solution of \eq{constraint}
with the correct Hermiticity properties, \CF~\cite{heneutr} reads:
\BEQ
\chi(z) =- \int dz'
 \bra{z} \Omega \left( B^{A}\right) \ket{ z'} 
 \frac{1}{\sqrt{2}}\gamma^{t}
\left( i\gamma^{A}\partial_{A} +m - i\mu \frac{1}{ \sqrt{2}} B_{3}\,
\epsilon_{AB}\gamma^{A}\gamma^{B}\right)\phi\left(z'\right)
\label{constraintsolved}
\EEQ
Here all arguments in which \eq{constraintsolved} is local, (\IE
$x^{A}$ and $t$) have been suppressed.
The matrix element of the operator $\Omega$ is given by:
\BEQ
\bra{ z} \Omega \left( B^{A}\right) \ket{ z'} 
 = \frac{i}{\sqrt{2}}\exp\left(\mu \sqrt{2} 
\int_{z'}^{z} dz''\epsilon_{AB}\gamma^{A}B^{B}\left(z''\right)\right)
\frac{1}{2}
\epsilon\left( z - z'\right)
\label{Omega}
\EEQ

Let us notice that solving the constraint eliminates two components of
the original, four component Dirac spinor. Therefore, instead of the
original Dirac matrices one can go over to $2\times 2$ Pauli matrices.
One easily verifies that $-i\epsilon^{AB}\gamma_{B}\rightarrow
\sigma^{A}$ gives the correct representation. We also introduce the
Hermitean operator, $p_{A} = -i \partial_{A}$ for the transverse
degrees of freedom.

It is now a matter of straightforward algebra to use \eq{constraintsolved}
in order to
eliminate the variable $\chi $ from the Lagrangian. We merely quote
the result; it can be conveniently written in Hamiltonian form as follows.
\begin{eqnarray}
L & = & \pi \partial_{t}\phi -H \nonumber \\ 
H & = & -2\mu \phi^{\dag}\sigma^{A}B_{A}\phi \nonumber \\
  & + & \phi^{\dag}\left( -i \sigma_{B}\epsilon^{BC}p_{C} +
m -\mu \sqrt{2}B_{3}\sigma_{3}\right)\nonumber \\ 
& \times & \Omega  \left( -B^{A}\right)\nonumber \\
 & \times & \left( i\sigma_{R}\epsilon^{RS}p_{S}
 + m -\mu \sqrt{2} B_{3}\sigma_{3}\right)\phi\nonumber \\
  &   &
\label{Hamiltonian}
\end{eqnarray}
The canonical momentum is given by $\pi = i \sqrt{2}\, \phi^{\dag}$.
(Of course, the odd looking factor of $\sqrt{2}$ in the 
definition of the canonical momentum can be eliminated by
rescaling the time variable.)

In \eq{Hamiltonian} we omitted all the symbols of integration over the
longitudinal coordinate, $z$; terms in which the operator $\Omega$
does not appear explicitly are local in all variables.

The Hamiltonian appearing in \eq{Hamiltonian} is  {\em exact\/.}
However, it is given by a rather complicated, non local  and
non linear expression:
this is the cost we have to pay for explicitly eliminating the 
constraint. We now argue that one can introduce physically reasonable
simplifications, as a result of which the problem becomes a manageable
one.  We notice that the exponential appearing in \eq{Omega}, regarded
as a matrix in the space of spinors is unitary, hence all matrix
elements are bounded.
 Therefore, one expects that at large values 
of $|z - z'|$ the exponent
oscillates rapidly and  contributes little to the Hamiltonian.
The dominant contribution is, probably, coming from small values of
 the difference of longitudinal coordinates. Hence, it appears  reasonable
to approximate the exponential in \eq{Omega} by 1. It is to be 
emphasized, however that the linearization of the effective Hamiltonian
in the magnetic field is not necessarily a harmless approximation.
Many instances are known when a similar truncation of a dynamical system
leads to the loss of important physics.  The validity
of this approximation therefore needs further study.

We now observe that, after having made the approximation just discussed,
there remain three types of terms in \eq{Hamiltonian}:
\begin{enumerate}
\item The local term, $\propto \sigma^{A}B_{A}$: this is purely a helicity 
flip term in terms of the lightlike helicity defined by $\sigma_{3}$.
\item Transverse spin--orbit coupling terms , $\propto \sigma_{A}
\epsilon^{AB}p_{B}$: these terms also give rise to the transverse
kinetic energy term, since $\sigma^{A}\epsilon_{AB}p^{B}\sigma_{R}
\epsilon^{RS}p_{S} = p_{A}p^{A}$.
\item ``Mass'' terms, $\propto M_{eff} = m - \mu \sqrt{2}\, 
\sigma_{3}B^{3}$.
In this picture, the longitudinal component of the magnetic field
plays the same role as the mass does. One has to recall in particular
that, in the case of Majorana neutrinos, the two helicity states do not 
necessarily have the same mass.
\end{enumerate}
\section{A Solvable Model}
\label{solvable}
We now turn to the description of a solvable model. Two further
approximations are made in order to accomplish this. {\em First}\/, we
consider a model with a reduced number of flavors: we suppress 
the flavor matrix structure in preceding equations altogether.
This means that in the case of Dirac neutrinos we are dealing with
a single flavor of two helicities, one corresponding to
an active, the other one to a sterile neutrino. In the case
of Majorana neutrinos we have genuinely two flavors. One hopes that, 
despite the reduction in the number of flavors, the qualitative
insight gained by solving the model  will facilitate the study
of more realistic models. 
{\em Second}\/,
we consider the high energy limit of the theory developed
so far: this is particularly easy in the front form of the
dynamics. 

We observe that the terms in items 2) and 3) in the preceding
Section are
proportional to $\epsilon\left( z- z'\right)$. Hence, in a Fourier
representation, \VIZ upon writing
\BEQ 
\phi \left( t, z, x^{A}\right) = \int dk \tilde{\phi }
 \left( t, k, x^{A} \right) \exp \left( -i kz \right)
\EEQ
and
\BEQ 
\epsilon (z) = \frac{{\cal P}}{2\pi i}\int \frac{dk}{k} \exp
\left( -ikz \right),
\EEQ
one finds that at high energies ($k\gg m$) the terms in items 2) and 3) are
negligible with respect the local term in item 1). (In the last equation
${\cal P}$ stands for the principal value.)

Thus, if one is interested in high energy neutrinos, the Hamiltonian can
be approximated by a local one, given by the first term in \eq{Hamiltonian}.

For this reason, it is convenient to work in a coordinate
representation for the density matrix. The equation of motion then
reads:
\begin{eqnarray}
-i \partial_{t}\bra{z,\vec{x}}\rho (t)\ket{z', \vec{x'}}& = &\mu \sqrt{2}
\vec{\sigma }\cdot\vec{B}\left(\vec{x},\frac{z-t}{\sqrt{2}}\right)\bra{z,\vec{x}}
\rho (t) \ket{z', \vec{x'}} \nonumber \\
 & - & \mu \sqrt{2}\bra{z ,\vec{x}}\rho
\left(t\right)\ket{z',\vec{x'}}\vec{\sigma}\cdot \vec{B}\left(\vec{x'},
\frac{z'-t}{\sqrt{2}}\right)
\label{eqofmotion}
\end{eqnarray}

In this equation, $\vec{x}$ stands for the transverse part of the
coordinate and $ \vec{\sigma}\cdot \vec{B}$ is the  two dimensional
scalar
product in transverse space. Of course,  the coordinate $x^{3}$ had to
be expressed by $ z$ and $t$; hence the $t$-dependence in the magnetic
field.

We choose the initial condition so as to describe a neutrino produced
at $\vec{x}=0$ and with a fixed value of $k$:
\BEQ
\bra{z, \vec{x}}\rho \left(0\right)\ket{z'\vec{x'}}
= \delta^{2}\left( \vec{x}\right)\delta^{2}\left( \vec{x'}\right)
\frac{\exp ik\left(z - z'\right)}{2\pi k}\rho_{s}\left(0\right),
\label{initialcondition} 
\EEQ
where $\rho_{s}(0)$ is the initial value of the spin density matrix.

The variable $k$ being large, the function
$\exp ik\left(z - z'\right)$ is  rapidly oscillating  unless
$z\approx z'$: it is permissible to put $z=z'$ in the
coefficient  of the exponential in \eq{initialcondition}. Further, in
the
approximation used,  the
dynamics described by eq.~\eq{eqofmotion}
is independent of $k$ and of $\vec{x}$.
Therefore, the dependence of $\rho (t)$ on these variables is
entirely
determined by the initial
condition and  the dynamical equation reduces to an
equation 
involving the spin density matrix alone,  as in non relativistic
spin dynamics. From now on, we  omit the the subscript $s$ and
we have:
\BEQ
-i\partial_{t}\rho \left( t \right) = \mu \sqrt{2}\left[ \vec{\sigma}\cdot
\vec{B}\left(\frac{z-t}{\sqrt{2}}\right), \rho \left( t\right)\right]
\label{spindynamics}
\EEQ
(Here and in what follows, $\vec{x}=0$ is understood.)

This equation can be solved by the standard time ordered series, \VIZ
\begin{eqnarray}
\rho\left(t\right)&=&\rho\left( 0\right)\nonumber \\
& +&i\mu \sqrt{2} \int_{0}^{t}dt'\left[\vec{\sigma}\cdot
\vec{B}\left(\frac{z-t'}{\sqrt{2}}\right)
,
\rho\left(0\right)
\right]\nonumber \\
 &+& \frac{\left( i\mu \sqrt{2}\right)^{2}}{2!}\int_{0}^{t}
dt'dt''T\left( \left[ \vec{\sigma}\cdot
\vec{B}\left(\frac{z-t'}{\sqrt{2}}\right),
\left[\vec{\sigma}\cdot \vec{B}\left(\frac{z-t''}{\sqrt{2}}\right),
\rho\left( 0 \right)\right]\right]\right)\nonumber \\ &+& \cdots
\label{timeordered}
\end{eqnarray}
We choose the initial condition as:
\BEQ \rho\left( 0\right) = \frac{1}{2}\left( 1 + S\sigma_{3}\right)
, \qquad \left(S^{2}\leq 1\right),
\EEQ
since neutrinos are produced with a definite helicity. (In  the case
of Dirac neutrinos, $S=\pm 1$, depending on whether a neutrino or
anti neutrino
is produced. In the case of Majorana neutrinos, $S$ may assume any
value between the limits stated above, depending on the production
mechanism.)

Next, we average the solution, \eq{timeordered} over the magnetic
field. We choose the generating functional of the moments as follows:
\begin{eqnarray}
Z[j] & = & \int {\cal D}B \exp -\left[ \frac{1}{2}\int
d^{3}xd^{3}x'B_{i}\left(x\right)C^{-1}_{ij}\left(x,x'\right)B_{j}
\left(x'\right)\right]\nonumber \\
& \times & \exp \int d^{3}x j_{i}
\left(x\right)B_{i}\left( x\right );\nonumber \\[1mm]
C_{ij}^{-1} & = & \frac{L}{4\pi \langle B^{2}\rangle}
\left(\delta_{ij}-\frac{\partial_{i}\partial_{j}}{\bigtriangledown^{2}}\right)
\left( L^{-2} - \bigtriangledown^{2}\right)^{2} \delta^{3}\left(x - x'\right).
\label{generatingfunctional}
\end{eqnarray} 
In the last equation, $L$ and $\langle B^{2}\rangle $ stand for the
correlation
length and mean square magnetic field, respectively. The measure is
normalized such that $Z\left[0\right]=1$.  
The transverse
projector is needed in order to make the correlation functions 
solenoidal. With the choice of the tensor $C^{-1}$ given in 
\eq{generatingfunctional}, the leading term in the long distance
behavior of the correlation function is $\propto \exp - \abs{x-x'}$.
In order to average equation \eq{timeordered} over the magnetic field,
one integrates over $B_{3}$ and sets the third component
of the source equal to zero. The  transverse generating functional
reads:
\begin{eqnarray}
Z_{T}&  = &\int {\cal D}\vec{B} \exp - \frac{L}{8\pi \langle B^{2}
\rangle } \int d^{3}x \left[ B^{A}(x) \left( \delta_{AB} - \frac{1}{2}
\frac{\partial_{A}\partial_{B}\left(x\right)}{\vec{\bigtriangledown}^{2}}\right)
B^{B}\right]\nonumber \\ 
& \times & \exp i \int d^{3} x \vec{j}\left(x\right)\cdot \vec{B}\left(x\right)
\label{transversegenerator}
\end{eqnarray}
  
We now notice that in the equation \eq{timeordered}, terms containing
odd
powers of $\mu$ are also odd in $B^{A}$. Therefore, in the limit
$\vec{j} \rightarrow 0$ the average of those terms vanishes. The even
terms in the series are obtained by taking the appropriate functional
derivatives of \eq{transversegenerator}. All of them are expressed in
terms of multiple time integrals of $C_{ij}\left(\abs{t-t'}\right)$
and its powers: those integrations are easily performed. It is
sufficient to illustrate the procedure for the second order term in 
\eq{timeordered}.

Carrying out the integrations, one gets:
\[
 - \mu^{2} \frac{1}{2} S\langle \int_{0}^{t} dt' dt'' \left[
\vec{\sigma}
\cdot \vec{B},\left[
\vec{\sigma}\cdot \vec{B}, \sigma_{3}\right]\right]\rangle
= - \mu^{2}\sigma_{3} \langle B^{2}\rangle tL \left( 1 - \exp
-\frac{t}{L}
\right)
\]
For large times the result in the last equation is
just proportional to $t$. The higher order terms follow a similar
pattern. The end result is:
\BEQ
\langle \rho \left( t \right) \rangle \sim
\frac{1}{2}\left( 1 + S\sigma_{3}\exp - \frac{t}{T}\right),
\label{asymptote}
\EEQ
with
\BEQ  \frac{1}{T} = 2\mu^{2} \langle B^{2}\rangle L .
\label{dampingcoefficient}
\EEQ
\section{Discussion}
\label{discussion}
The solvable model described in the last section leads to
the remarkable result that 
in the random field
the behavior of the helicities is an {\em ergodic} one: irrespective
of what the initial density matrix was, for $t\gg T$, the helicities
are equally distributed. In the case of Dirac neutrinos, this is 
rather uninteresting: roughly 1/2 of them is a sterile
one. Unfortunately,
however,
calculated neutrino fluxes emerging from such astrophysical objects
as an AGN usually cannot be trusted to an accuracy which would
permit an observational testing of the result. In the case of 
Majorana neutrinos, however, the two helicity states correspond to 
two different flavors. Given the fact that the neutrinos produced
arise mostly from pion decay, the presence of the other flavor in
roughly equal proportion is an observationally testable result.
We conjecture that the situation is similar if all neutrino flavors
are properly taken into account. Should this conjecture be verified by
future calculations, one would have a very important observational
prediction. If neutrinos are Majorana particles and they 
possess substantial (say, $10^{-12}\mu_{B}$)
transition moments, the flavor distribution of neutrino events observed 
in a high energy neutrino telescope would be practically uniform.
For this reason it would be of utmost importance to
\begin{itemize}
\item observe high energy neutrinos emerging from point sources,
\item develop flavor sensitive detection techniques in neutrino
telescopes.
\end{itemize}

We remark that the question of flavor change in the case of
magnetized Majorana neutrinos can be investigated in terrestrial
neutrino experiments as well, see~\cite{primakoff}. The two types of
experiments referred to above are complementary to each other. If
both types indicate the presence of flavor flipping, one would have
a window on post standard model physics as well as the structure
of important point sources of high energy neutrinos.

From the theoretical  point of view, the main result of this
work is the development of a consistent formalism in order to
treat the propagation of high energy neutrinos in a random
magnetic field (more generally, in a random environment). 
This problem has been treated before, see for instance,
ref.~\cite{elmforsetal} and the virtually complete bibliography
quoted there. (The authors quoted in those references concentrate
mostly on the early universe and the role of sterile 
neutrinos in it\footnote{In most of the works quoted, an average is
taken at the level of the
evolution equations rather than their solutions. It has been
known for some time that  averaging
the evolution equations 
over a random variable  is an incomplete procedure
(important physics may be  missed). It may also lead  to internal 
contradictions. For a lucid exposition, see \cite{vankampen}.
A notable exception is ref.~ \cite{nicolaidis}, where a 
simplified equation is given a correct treatment.}
The main advantage of the formalism developed 
in refs.~\cite{martin, dedominicis} and adapted 
to highly relativistic systems 
 in this paper is twofold: it is guaranteed to be free of
internal inconsistencies and it allows the development of controllable
approximation methods.

A final comment is in order. Assuming that one
observes flavor conversion in some AGN, one may ask whether the
theory developed here can be used to test some properties of the magnetic
fields in the source. We  argue that the answer is in the
affirmative to some extent. The measurable quantity characterizing
the magnetic field is the coefficient given by
equation~\eq{dampingcoefficient}. One notices that this expression
contains the quantities characterizing the magnetic field 
only in the combination $\mu^2\langle B \rangle^{2}L$. 
This is not an accident. There is an important  class of
distributions of the magnetic field characterized by the following: 
\begin{enumerate}
\item  The magnetic field has a zero mean value,
\item The distribution  is approximately Gaussian 
(higher cumulants are approximately zero), 
\item  the distribution contains only one length scale. 
\end{enumerate} 
Every such distribution  leads to a damping
coefficient of a form which can differ from \eq{dampingcoefficient}
only in a numerical factor. In essence, this follows from
dimensional analysis. 

In fact, the magnetic moment and the magnetic
field strength enter the Hamiltonian in the combination $\mu B$;
since $\langle B\rangle=0$, this combination must enter 
a physical observable quadratically.
Finally, in order to form a quantity of inverse length 
(the damping coefficient), the expression
has to be multiplied by a characteristic length in the distribution.

As another  example,  Enquist \ETAL, \cite{enquist} derived a damping
coefficient of this form, even though they use a field distribution
quite different form ours\footnote{We thank Kari Enquist for
pointing out the similarity of the damping coefficients in
refs.\cite{pletter} and \cite{enquist}.}. (The  distribution in
\cite{enquist} has a vanishing correlation length, but a finite domain
size.) Thus measuring such quantities as the  damping
coefficient can only yield information about the class of the magnetic
field distribution, not about its details.
\begin{flushleft}
{\large\bf Acknowledgement}
\end{flushleft}
This work was performed during the authors' visit at the Dipartimento
di Fisica, Universit\'{a} di Firenze. We wish to thank Roberto
Casalbuoni, Director of the Department for the hospitality extended to
us. We thank Bianca Monteleoni for useful conversations on
observational neutrino astrophysics and Kari Enqvist for
some useful critical remarks.

\end{document}